\documentclass[aps,eqsecnum,twocolumn,prd,showpacs,floatfix]{revtex4}
\usepackage{graphicx}
\def\bold#1{\setbox0=\hbox{$#1$}%
     \kern-.025em\copy0\kern-\wd0
     \kern.05em\%\baselineskip=18ptemptcopy0\kern-\wd0
     \kern-.025em\raise.0433em\box0 }
\def\slash#1{\setbox0=\hbox{$#1$}#1\hskip-\wd0\dimen0=5pt\advance
         to\wd0{\hss\sl/\/\hss}}

\newcommand{\be}{\begin{equation}}
\newcommand{\ee}{\end{equation}}
\newcommand{\bea}{\begin{eqnarray}}
\newcommand{\eea}{\end{eqnarray}}
\newcommand{\nn}{\nonumber}
\newcommand{\dd}{\displaystyle}

\newcommand{\spur}[1]{\not\! #1 \,}

\begin{document}
\begin{titlepage}
\addtolength{\jot}{10pt}

 \preprint{\vbox{\hbox{BARI-TH/07-572 \hfill}
                \hbox{September 2007\hfill} }}

\title{\bf  FCNC $B_s$ and $\Lambda_b$ transitions: \\
\vspace*{0.3cm}
Standard Model versus a single Universal Extra Dimension  scenario\\
\vspace*{0.3cm}
}

\author{P. Colangelo$^a$, F. De Fazio$^a$, R. Ferrandes$^{a,b}$,  T.N. Pham$^c$ \\}

\affiliation{ $^a$ Istituto Nazionale di Fisica Nucleare, Sezione di Bari, Italy\\
$^b$ Dipartimento di Fisica,  Universit\'a  di Bari, Italy\\
$^c$ Centre de Physique Th\'eorique, 
\'Ecole Polytechnique, CNRS, 91128 Palaiseau, France \\}

\begin{abstract}
We study the FCNC $B_s \to \phi \gamma, \phi \nu \bar \nu$ and $\Lambda_b  \to \Lambda \gamma, \Lambda \nu \bar \nu$ transitions 
in the Standard Model and in a scenario with a single Universal Extra Dimension. In particular, we focus on the present knowledge of the hadronic uncertainties and on  possible improvements. We discuss how the measurements of these modes can be used to
constrain the  new parameter involved in the extra dimensional scenario, the radius $R$ of the extra dimension, 
completing the information  available from  B-factories.  The rates of these $b \to s$ induced decays are within the reach of  new experiments,  such as LHCb.   

 \end{abstract}

\vspace*{1cm} 

\maketitle
\end{titlepage}

\newpage
\section{Introduction}\label{sec:intro}
The heavy flavour Physics programmes at the hadron facilities, the Tevatron at Fermilab  and the  
Large Hadron Collider (LHC) at CERN, include the analysis of heavy particles, in particular $B_s$ and 
$\Lambda_b$, which cannot be produced  at the $e^+ e^-$ factories operating at the peak of 
$\Upsilon(4S)$ \cite{Ball:2000ba}.
These programmes involve as an important topic  the study of  processes induced by  Flavour Changing Neutral Current
(FCNC)  $b \to s$ transitions, since they  provide us with   tests  of  the Standard Model (SM) and constraints for  New Physics (NP) scenarios. 
 The observation  of  $B_s^0 - \overline{ B_s^0}$ oscillations at the Tevatron  \cite{mixingbs}
represents  a great success in this Physics programme.

Considering the experimental situation, data are available at present
for several  $b \to s$  FCNC  $B$ meson decays.
Together with the inclusive radiative $B \to X_s \gamma$ branching ratio, the rates of a few
exclusive radiative modes, both for charged $B$:
  $B^\pm \to K^{*\pm} \gamma$,  $K_1(1270)^\pm \gamma$
 and $K^*_2(1430)^\pm \gamma$, both for neutral $B$:
 $B^0 \to K^{*0} \gamma$ 
 and $K^*_2(1430)^0 \gamma$ have been measured  \cite{PDG}.
 Moreover, the branching fractions of
 $B^\pm \to K^{(*)\pm} \ell^+ \ell^-$ and   $B^0 \to K^{(*)0} \ell^+ \ell^-$, with $\ell=e,\,\mu$ have been determined by  the Belle and BaBar collaborations, which   have also
provided  us with preliminary measurements of  the lepton
forward-backward asymmetry
\cite{Abe:2004ir,Aubert:2005cf}  together with   the $K^*$ longitudinal
helicity fraction in  $B \to K^* \ell^+ \ell^-$ \cite{Aubert:2006vb}.  Furthermore, 
upper bounds for
  ${\cal B}(B \to K^{(*)} \nu \bar \nu)$  have  been established  \cite{PDG}. 
  Even not considering non leptonic $b \to s$ penguin induced  $B$ decays, the interpretation
  of which is not straightforward,  this wealth of 
measurements has  already severely constrained the parameter space
of various non standard scenarios.  Other tight  information  could be
obtained by  more precise data in the modes already observed, as well as by  the 
measurement of the branching fractions and of the spectra of  $B
\to K^{(*)} \nu \bar \nu$,  and  of  $B \to K^{(*)} \tau^+ \tau^-$ in which  
 the $\tau$ polarization asymmetries are    sensitive to Physics beyond  SM.

Other processes induced by $b \to s$ transition involve  the hadrons $B_s$, $B_c$ and $\Lambda_b$,  the
decay modes of which are  harder to be experimentally studied, e.g.  due to
their  smaller production rate in b quark hadronization  with respect to $B$ mesons. However, at the hadron colliders, in particular at LHC, the 
number of produced particles is  so large that even these processes are expected to be observable,
so that their study can contribute to our  understanding of Physics of rare transitions within and
beyond the Standard Model description of elementary interactions.

In this paper  we study a few $b \to s$ FCNC induced $B_s$ and $\Lambda_b$
decays, in particular $B_s \to \phi \gamma, \phi \nu \bar \nu$ and $\Lambda_b  \to \Lambda \gamma, \Lambda \nu \bar \nu$, in SM and   in  a New Physics scenario where
 a single universal extra dimension is considered,  the  Appelquist,
Cheng and Dobrescu (ACD) model
\cite{Appelquist:2000nn}.
Models with extra dimensions have been proposed  as viable
 candidates to solve some   problems affecting  SM \cite{rev}, and within this class of NP models the  ACD scenario with a single extra dimension is worth investigating due to its appealing features. 
Here, we do not discuss in details the various aspects of the model,  which have been
worked out in \cite{Appelquist:2000nn} and are summarized in
\cite{Buras:2002ej}-\cite{noi}.  We only recall  that
the model consists in a minimal extension of SM in $4+1$ dimensions,  with the extra dimension
compactified to the orbifold $S^1/Z_2$.
The fifth coordinate $y$ runs from $0$ to $2 \pi R$, with
 $y=0$, $y=\pi R$  fixed points of the orbifold,  and $R$ the radius of the orbifold 
 which represents a new physical parameter. All the fields  are allowed to propagate
 in all  dimensions, therefore the model belongs to the class of {\it universal} extra dimension scenarios.
 The four-dimensional description  includes SM particles,
corresponding to the zero modes of fields propagating in the compactified extra dimension,
together with  towers of  Kaluza-Klein (KK) excitations corresponding to the higher modes. Such
 fields  are imposed to be even 
under parity transformation in the fifth coordinate $P_5: y \to -y$. Fields  which are odd under $P_5$
propagate in the extra dimension without zero modes and correspond to particles  with
 no SM partners. 
  
In addition to 5-d bulk terms, the Lagrangian of the ACD model may also include boundary terms which represent additional parameters of the theory and get renormalized by bulk interactions, although
being volume suppressed. A simplifying assumption is that such boundary terms vanish at the cut-off scale, so that  a minimal Universal Extra Dimension model can be defined  in which the only new parameter with respect to  the Standard Model is the radius $R$ of the extra dimension, an important feature as far as the phenomenological investigation of the model is concerned \cite{georgi}. 

The masses of  KK particles  depend on  the radius $R$ of the extra dimension.  For example,
  the  masses of the KK bosonic modes are given by
 \cite{Appelquist:2000nn}, \cite{Buras:2002ej}-\cite{noi}:  
 \be m_n^2=m_0^2+{n^2 \over R^2} \,\,\,\,\,\,\, n=1,2,\dots \ee 
For small values of $R$
these particles, being more and more massive, decouple from the
low energy regime. Thus,   within this  model   candidates for dark matter are available,  the 
Kaluza-Klein (KK) excitations of the photon or of the neutrinos
with KK number $n=1$ \cite{Cheng:2002iz,Hooper:2007qk}. This is  related to
another  property of the ACD model,  the conservation    of the KK parity $(-1)^j$, with $j$  the KK number 
\cite{footnote}.
KK parity conservation implies  the absence of tree level contributions of Kaluza Klein states to
processes taking place at low energy, $\mu \ll 1/R$, a feature which  permits to establish  a
 bound:  ${1 / R} \ge 250-300$ GeV by the analysis of 
Tevatron run I data  \cite{Appelquist:2002wb}.  Moreover, the precision measurements
of electroweak observables at the LEP collider allow to obtain the bound ${1 / R} \ge 600$ GeV
assuming for the Higgs a mass of 115 GeV, a constraint  on the extra dimension radius which can be relaxed as low as ${1 / R} \ge 300$ GeV with increasing Higgs mass \cite{Gogoladze:2006br}.
Other bounds can be derived invoking  cosmological arguments;   in 
Ref.\cite{Cembranos:2006gt}  it was found that the region of parameters preferred by cosmological constraints in a single UED scenario corresponds to a Higgs mass between 185 and  245 GeV, to a lightest KK particle between 
810 and 1400 GeV and to  maximal spitting between the first KK modes of 320 GeV;  such constraints
come from the diffuse photon spectrum and from the null searches of exotic stable charged particles,  
 assuming that  the first excitation of the graviton is the lightest  KK
particle.
 These bounds, however, are different if the lightest KK particle is not the first excitation of the graviton.

The fact that KK excitations can influence processes occurring at loop level suggests that FCNC transitions are particularly suitable for constraining the extra dimension model, providing us with many
observables sensitive to the compactification radius $R$. For this reason,  
in \cite{Buras:2002ej,Buras:2003mk} the effective Hamiltonian
governing $b \to s$ transitions was  derived,  and  inclusive $B \to
X_s \gamma$, $B \to X_s \ell^+ \ell^-$, $B \to X_s \nu \bar \nu$
transitions, together with the  $B_{s (d)}$ mixing,  were studied. In
particular, it was found that   ${\cal B}(B \to X_s
\gamma)$ allowed to constrain ${1/ R} \ge 250$
GeV, a bound updated by  a more recent analysis based on   
 the NNLO value of the SM Wilson coefficient $c_7$ (defined in the next section) \cite{NNLO} and on  new experimental data    to $1/R \ge 600$ GeV at 95$\%$ CL,
or to  $1/R \ge 330 $ GeV at 99$\%$ CL  \cite{Haisch:2007vb}.
Concerning the   exclusive modes $B \to K^* \gamma$, $B \to
K^{(*)} \ell^+ \ell^-$, $B \to K^{(*)} \nu \bar \nu$ and
 $B \to K^{(*)} \tau^+ \tau^-$,   it was argued
 that the uncertainty related to the hadronic
matrix elements does not obscure the sensitivity to the
compactification parameter $R$, and that current data, in particular
the decay rates of $B \to K^* \gamma$ and $B \to K^* \ell^+
\ell^-$ $(\ell=e,\mu)$ can provide the  bound  $1/R \geq 300-400$
GeV  \cite{noi,Colangelo:2006gv}.

In case of $B_s$,  the mode $B_s \to \mu^+ \mu^-$ was
recognized as the process receiving the largest enhancement with
respect to the SM prediction;  however,  since the   predicted branching
fraction is  ${\cal O}(10^{-9})$,   the
 measurement  is very challenging   \cite{Buras:2002ej}. 
 A modest enhancement  was   found in $B_s \to \gamma
\gamma$ \cite{Devidze:2005ua}, in  $B_s \to \phi \ell^+ \ell^-$ and in 
$B_s \to \ell^+ \ell^- \gamma$ \cite{Mohanta:2006ae}.
 In the case of $\Lambda_b$, for $1/R \simeq 300$ GeV
 a  sizeable effect of the extra dimension
 was found in  $\Lambda_b \to \Lambda \ell^+ \ell^-$  \cite{Aliev:2006xd}, analogously to what obtained
 in  $B \to K^{(*)} \ell^+ \ell^-$.

For  the   modes  $B_s \to \phi \nu \bar \nu$,  $\Lambda_b \to \Lambda
\gamma$ and  $\Lambda_b \to \Lambda \nu \bar \nu$ analyzed in this paper,  no  data
are  available at present, while for $B_s\to\phi \gamma$ a first  measurement of the branching fraction has been recently carried out.
  Our aim is to work out a set of predictions for various
observables in the Standard Model,  making use also of information obtained at the $B$ factories. Moreover, we consider these processes in the extra dimension scenario studying
 the dependence on $R$  which could provide us with  further ways  to constrain such a  parameter. To be conservative,  we  consider  the   range of $1/R$ starting from ${1 / R} \ge 200$ GeV. In the lowest
part of this range   the effect of the extra dimension is clearly visible in the various observables we  consider;
 obviously, the bounds previously mentioned  must be taken into account in the final considerations.
 
In the next Section we recall the effective Hamiltonian inducing
$b \to s \gamma$ and $b \to s \nu \bar \nu$ decays in  SM and
in the ACD model. The case of $B_s$ is considered in Section
\ref{sec:bs}, while the  $\Lambda_b$ transitions are the
subject of Section \ref{sec:lambdab}. The last Section is devoted
to the conclusions.

\section{ $b \to s \gamma$ and $b \to s \nu \bar \nu$ Effective Hamiltonians} \label{sec:hamiltonian}

In the Standard Model   the  $b \to s \gamma$ and  $b
\to s \nu \bar \nu$  transitions are described by the effective $\Delta B=-1$,
$\Delta S=1$ Hamiltonians
 \begin{equation} H_{b \to s \gamma}=4\,{G_F
\over \sqrt{2}} V_{tb} V_{ts}^*  c_7^{eff}(\mu) O_7(\mu) \label{hamilgamma} 
\end{equation}
and
 \begin{equation}
H_{b \to s\nu \bar \nu}= {G_F \over \sqrt{2}} {\alpha (M_W)
\over 2 \pi \sin^2(\theta_W)} V_{tb} V_{ts}^* \eta_X X(x_t) \, O_L  = c_L O_L\label{hamilnu}
\end{equation}
involving the operators
 \begin{equation}
O_7 = {e \over 16 \pi^2} \big[ m_b ({\bar s}_{L }
\sigma^{\mu \nu} b_{R }) +m_s ({\bar s}_{R }
\sigma^{\mu \nu} b_{L }) \big]F_{\mu \nu} \label{opgamma} 
\end{equation}
and 
 \begin{equation}
O_L = {\bar s}\gamma^\mu (1-\gamma_5) b {\bar \nu}\gamma_\mu
(1-\gamma_5) \nu \,\,,\label{opnu}
\end{equation}
respectively.  Eq.(\ref{hamilgamma}) describes
 magnetic penguin diagrams, while (\ref{hamilnu}) 
is obtained from $Z^0$ penguin and box diagrams,  with  the
dominant contribution corresponding to a top quark intermediate
state. $G_F$ is the Fermi constant and $V_{ij}$ are elements of the CKM mixing matrix; moreover,
$\displaystyle
b_{R,L}={1 \pm \gamma_5 \over 2}b$, $\alpha$ is the electromagnetic constant, $\theta_W$  the
Weinberg angle and $F_{\mu \nu}$ denotes the electromagnetic field
strength tensor. The  function $X(x_t)$ ($x_t=\displaystyle{ m_t^2
\over M_W^2}$,  with $m_t$  the top quark mass) has been computed
in \cite{inami} and \cite{buchalla}; 
the QCD factor $\eta_X$ is close to one, so that we can put  it to unity \cite{urban,Buchalla:1998ba}.

In eq. (\ref{hamilgamma}) we use an effective coefficient $c_7^{eff}$ which turns out to be scheme
independent and takes into account the mixing between the  operators $O_8$ with $O_7$  under renormalization group evolution \cite{Buras:1993xp}.

In the ACD model  no  operators other than those in 
(\ref{hamilgamma}) and (\ref{hamilnu}) 
 contribute to $b \to s \gamma$ and  $b \to s \nu \bar \nu $ transitions:
  the model  belongs to the class of Minimal Flavour Violating models,
 where  effects beyond SM are only encoded in the Wilson coefficients of the effective Hamiltonian   \cite{Buras:2002ej,Buras:2003mk}.
   KK excitations  only  modify  $c_7$ and $c_L$ inducing a
dependence on the  compactification radius $R$.  For
large values of $1/R$, due to decoupling of  massive KK states,
    the Wilson coefficients reproduce the  Standard Model values,  so that the SM phenomenology is  recovered.  As a general expression, the Wilson
coefficients are represented by functions
$F(x_t,1/R )$ generalizing  their SM analogues
$F_0(x_t)$: \be F(x_t,1/R)=F_0(x_t)+\sum_{n=1}^\infty
F_n(x_t,x_n) \,, \label{fxt} \ee
 with
$x_n=\displaystyle{ m_n^2 \over M_W^2}$ and  $m_n=\displaystyle{n
\over R}$.  
A remarkable result is that the sum over the KK contributions in  (\ref{fxt}) is finite  at the leading order (LO)  in all cases as a
consequence of a generalized GIM mechanism  \cite{Buras:2002ej,Buras:2003mk}. When  $R \to 0$ the Standard Model result is obtained  since 
$F(x_t,1/R) \to F_0(x_t)$ in that limit.

For $1/R$  of the order of a few hundreds of GeV the
coefficients differ from their Standard Model value; in particular
 $c_7$ is suppressed, as one can infer considering the  expressions 
  collected in  \cite{Buras:2002ej}-\cite{noi} together with the function $X(x_t,1/R)$.
Therefore, the predicted widths and spectra are modified with respect to SM.   
In case of exclusive decays, it is important to study if this  effect is obscured  by 
the hadronic uncertainties,  a discussion that we  present in the following two Sections for
$B_s$ and $\Lambda_b$, respectively.

\section{$B_s \to\phi \gamma$ and $B_s \to \phi \nu \bar \nu$ decays}\label{sec:bs}

The description of the decay modes $B_s \to \phi \gamma$ and  $B_s \to \phi
\nu \bar \nu$  involves the hadronic matrix elements of the operators
appearing in the effective Hamiltonians
(\ref{hamilgamma})-(\ref{hamilnu}). In case of  $B_s \to \phi \gamma$
 the matrix element of $O_7$ can be parameterized in terms of three form factors:
\begin{eqnarray}
&&<\phi(p^\prime,\epsilon)|{\bar s} \sigma_{\mu \nu} q^\nu
{(1+\gamma_5) \over 2} b |B_s(p)> \nonumber \\&=& i \epsilon_{\mu
\nu \alpha \beta} \epsilon^{* \nu} p^\alpha p^{\prime \beta}
\; 2 \; T_1(q^2)  + \nonumber \\
&+&  \Big[ \epsilon^*_\mu (M_{B_s}^2 - M^2_\phi)  -
(\epsilon^* \cdot q) (p+p')_\mu \Big] \; T_2(q^2) \nonumber \\
&+& (\epsilon^* \cdot q) \left [ q_\mu - {q^2 \over M_{B_s}^2 -
M^2_\phi} (p + p')_\mu \right ] \; T_3(q^2)  \; , \nonumber \\
\label{t1}
\end{eqnarray}
\noindent 
where $q=p-p^\prime$ is the momentum of  the photon  and $\epsilon$  the $\phi$ meson polarization vector.  At zero value of $q^2$ the condition  $T_1(0)=T_2(0)$ holds, so that the 
$B_s \to \phi \gamma$ decay amplitude involves a single hadronic parameter,  $T_1(0)$.
On the other hand, the matrix
element of $O_L$ can be parameterized as follows:
\begin{eqnarray}
&&<\phi(p^\prime,\epsilon)|{\bar s} \gamma_\mu (1-\gamma_5) b
|B_s(p)> \nonumber \\&=& \epsilon_{\mu \nu \alpha \beta}
\epsilon^{* \nu} p^\alpha p^{\prime \beta}
{ 2 V(q^2) \over M_{B_s} + M_\phi} 
- i \Big [ \epsilon^*_\mu (M_{B_s} + M_\phi) A_1(q^2)   \nonumber \\
&-&
(\epsilon^* \cdot q) (p+p')_\mu  {A_2(q^2) \over M_{B_s} +
M_\phi } \nonumber \\
&-&  (\epsilon^* \cdot q) {2 M_\phi \over q^2} \big(A_3(q^2)
- A_0(q^2)\big) q_\mu \Big] \label{a1}
\end{eqnarray}
\noindent with a relation holding among the form factors $A_1$, $A_2$ and $A_3$:
\begin{equation}
A_3(q^2) = {M_{B_s} + M_\phi \over 2 M_\phi}  A_1(q^2) - {M_{B_s}
- M_\phi \over 2 M_\phi}  A_2(q^2)
\end{equation}
together with  $ A_3(0) = A_0(0)$.

The  form factors represent a source of uncertainty in  predicting the
$B_s$  decay rates we are considering.  The other parameters are 
 fixed to  $m_b=4.8 \pm 0.2$ GeV and  $M_{B_s}=5.3675\pm0.0018$ GeV,
 together with
$\tau_{B^0_s}=(1.466\pm0.052) \times 10^{-12}$ s, $V_{tb}=0.999$ and $V_{ts}=0.0406\pm0.0027$,
using the central values and the uncertainties  quoted by the Particle Data Group \cite{PDG}.

 \subsection{$B_s \to\phi \gamma$}

\begin{figure}[t]
\begin{center}
\includegraphics[width=0.40\textwidth] {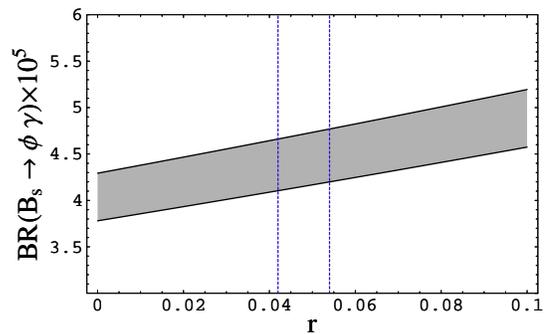} \\
\end{center}
\caption{\baselineskip=12pt  ${\cal B}(B_s \to \phi \gamma)$ as a function of the $SU(3)_F$ breaking parameter $r$
in the ratio of $B_s \to \phi$ vs  $B_d \to K^*$ form factors $T_1(0)$ (\ref{ratior}).  The  dashed vertical lines show the range of   $r$  obtained by   LCSR.} 
\vspace*{1.0cm} \label{fig:bsphigamma0}
\end{figure}

From the effective Hamiltonian  (\ref{hamilgamma}),  together with (\ref{opgamma})  and the matrix element  (\ref{t1}),  it is straightforward to calculate the expression of the
$B_s \to\phi \gamma$ decay rate: 
 \bea \Gamma(B_s \to \phi \gamma)&=&{\alpha(0)
G_F^2 \over 8 \pi^4}|V_{tb}V_{ts}^*|^2 (m_b^2  + m_s^2) |c_7^{eff}|^2 [T_1(0)]^2
\nonumber \\ &\times & M_{B_s}^3 \left( 1-{M_\phi^2 \over
M_{B_s}^2} \right)^3 \,\,\, . \label{ratebsphigamma} \eea
This expression is useful to relate  the branching fraction ${\cal B}(B_s \to \phi \gamma)$ to the measured value of ${\cal B}(B_d \to K^{*0} \gamma$), since
\bea
&& {\cal B}(B_s \to \phi \gamma)= \left ({T_1^{B_s \to \phi}(0) \over T_1^{B_d \to K^{*0}}(0)}\right)^2 
\left( {M_{B_d} \over M_{B_s}} \right)^3 \hspace*{1cm}\nn \\
 && \times \left( {M_{B_s}^2-M_\phi^2 \over M_{B_d}^2-M_{K^{*0}}^2}\right)^3 {\tau_{B_s} \over  \tau_{B_d}} {\cal B}(B_d \to K^{*0} \gamma) \,\,\, \label{Bsratio}
\eea
where we have indicated  which hadronic matrix elements the form factors refer to. 
Eq.(\ref{Bsratio}) shows that, in addition to measured quantities, the crucial quantity to predict
${\cal B}(B_s \to \phi \gamma)$ is the $SU(3)_F$ breaking parameter $r$ defined by
\be
{T_1^{B_s \to \phi}(0) \over T_1^{B_d \to K^{*0}}(0)}=1+r  \,\,\, , \label{ratior}
\ee as shown in Fig.  \ref{fig:bsphigamma0}
for positive values of $r$,  using ${\cal B}(B_d \to K^{*0} \gamma)=(4.1 \pm 0.2) \times 10^{-5}$ and
 $\tau_{B_d}=(1.530\pm0.009) \times 10^{-12}$ s \cite{PDG}, and combining in quadrature the uncertainties of the various quantites in (\ref{Bsratio}).
 Detailed analyses of the range of values within which $r$ can vary
are not available, yet. Using $r=0.048\pm0.006$ estimated by Light Cone Sum Rules
(LCSR)  \cite{Ball:2004rg} we obtain the range bounded by the  dashed vertical lines in  Fig.  \ref{fig:bsphigamma0},  which allows us  to  predict: 
\be
{\cal B}(B_s \to \phi \gamma)=(4.2 \pm 0.3) \times 10^{-5} \,\, .
\ee 
%
%
Notice  that the chosen  value of $r$ is   smaller than an analogous quantity parameterizing the ratio of  leptonic constants    $f_{B_s}/ f_{B_d}$,  estimated as:
$r=0.09\pm0.03$ \cite{Blasi:1993fi}.

A compatible value  of  ${\cal B}(B_s \to \phi \gamma)$ is obtained in  SM  using the form factor 
$T_1(0)=(17.45\pm1.65)\times 10^{-2}$ determined by LCSR  \cite{Ball:2004rg}, although with a larger uncertainty:
 \be 
{\cal B}(B_s \to \phi \gamma)=(4.1 \pm 1.0) \times 10^{-5}
\,.\label{brbsgammasm}\ee
These results  must be  compared to  the   measurement:
\be
{\cal B}(B_s \to \phi \gamma)_{exp}=(5.7^ {+1.8}_{-1.5}(stat)^{+1.2}_{-1.1}(syst))\times 10^{-5}  \ee
recently carried out by Belle Collaboration in a run at the $\Upsilon(5S)$ peak
 \cite{Wicht:2007ni}; this experimental
result is  affected by a  statistical and systematical uncertainty which are expected to be reduced in the near future.

\begin{figure}[t]
\begin{center}
\vspace*{0.2cm}
\includegraphics[width=0.40\textwidth] {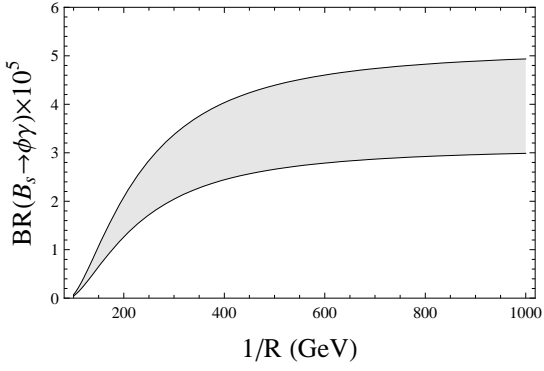} \\
\end{center}
\caption{\baselineskip=12pt  ${\cal B}(B_s \to \phi \gamma)$ {\it vs}  the inverse radius $1/R$
 of the compactified extra dimension in the ACD model. The Belle  measurement is:
 ${\cal B}(B_s \to \phi \gamma)_{exp}=(5.7^ {+1.8}_{-1.5}(stat)^{+1.2}_{-1.1}(syst))\times 10^{-5}$). 
} \vspace*{1.0cm} \label{fig:bsphigamma}
\end{figure}

In the single extra dimension scenario the modification of the Wilson coefficient $c_7^{eff}$,  corresponding to the variation of  the compactification radius $R$,  
changes  the  prediction for  ${\cal B}(B_s \to \phi \gamma)$, as shown  in
Fig. \ref{fig:bsphigamma}  where we plot the  branching ratio versus  $1/R$.  
The width of the band reflects the uncertainty on the form factor quoted above, as well as on the
parameters $m_b$ and $V_{ts}$. As for  $c_7^{eff}$, it is affected by the uncertainty due to the higher order corrections in the ACD model. The inclusion of next-to-leading order QCD corrections would require the contribution of two-loop diagrams involving KK gluon corrections which at present
is not known. One could estimate the size of this correction by considering the effect of the variation of the matching scale, as done  in \cite{Haisch:2007vb},  where it was found that changing this scale between  80 and 320 GeV  the value of ${\cal B}(B \to X_s \gamma)$ is affected by an uncertainty 
not exceeding  $|^{+8}_{-9}$ $\%$ for $1/R$ in the range between $200$ and $1500$ GeV. For the exclusive modes   considered here,  such an uncertainty,  expected to be similar, must be combined
in quadrature with the other errors, in particular with the hadronic uncertainty which is  at present the
largest one.

We observe that for low values of $1/R$  the branching ratio is reduced with respect to the SM expectation:   at, e.g.,  $1/R=300$ GeV    ${\cal B}(B_s \to \phi \gamma)$ is smaller by $35\%$ 
 as a consequence of the lower value of $c_7^{eff}$ in the ACD model. This effect was already noticed 
in the analysis of $B \to X_s \gamma$ and $B \to K^* \gamma$. For higher values of $1/R$  the lowering  of  the branching fraction is   obscured by  the hadronic uncertainty due to the form factor.

\subsection{$B_s \to\phi \nu \bar \nu$}

Analogously to the case of $B \to K^* \nu \bar \nu$ \cite{Colangelo:1996ay}, 
for this decay mode it is convenient to  separately consider the missing energy
distributions for  longitudinally (L) and transversely ($\pm$) polarized $\phi$ mesons: 
\bea {d \Gamma_L \over dx} &=& 3\,{ |c_L |^2 \over 24
\pi^3} { |\vec p^{~\prime}| \over M_\phi^2} \big[ (M_{B_s} +
M_\phi)(M_{B_s} E^\prime-M_\phi^2) A_1(q^2) \nonumber \\ &-& {2
M_{B_s}^2 \over M_{B_s} + M_\phi} |\vec p^{~\prime}|^2 A_2(q^2)
\big]^2\,, \label{long} \eea and \bea {d \Gamma_{\pm} \over dx}
&=& 3 { |\vec p^{~\prime}| q^2 \over 24 \pi^3} |c_L |^2 \big|  { 2
M_{B_s} |\vec p^{~\prime}| \over M_{B_s} + M_\phi} V(q^2)
\nonumber \\ &\mp& (M_{B_s} + M_\phi) A_1(q^2) \big|^2 \,\,\,\, .
\label{tran} \eea 
In (\ref{long})-(\ref{tran})  $x=E_{miss}/M_{B_s}$, with $E_{miss}$  the
energy of the neutrino pair (missing energy);   $q$ is the momentum
transferred to the neutrino pair,  $\vec p^{~\prime}$ and
$E^\prime$  the $\phi$ three-momentum and energy in the ${B_s}$
meson rest frame, and   the sum over the three neutrino species has been carried out.

\begin{figure}[t]
\begin{center}
 \includegraphics[width=0.37\textwidth] {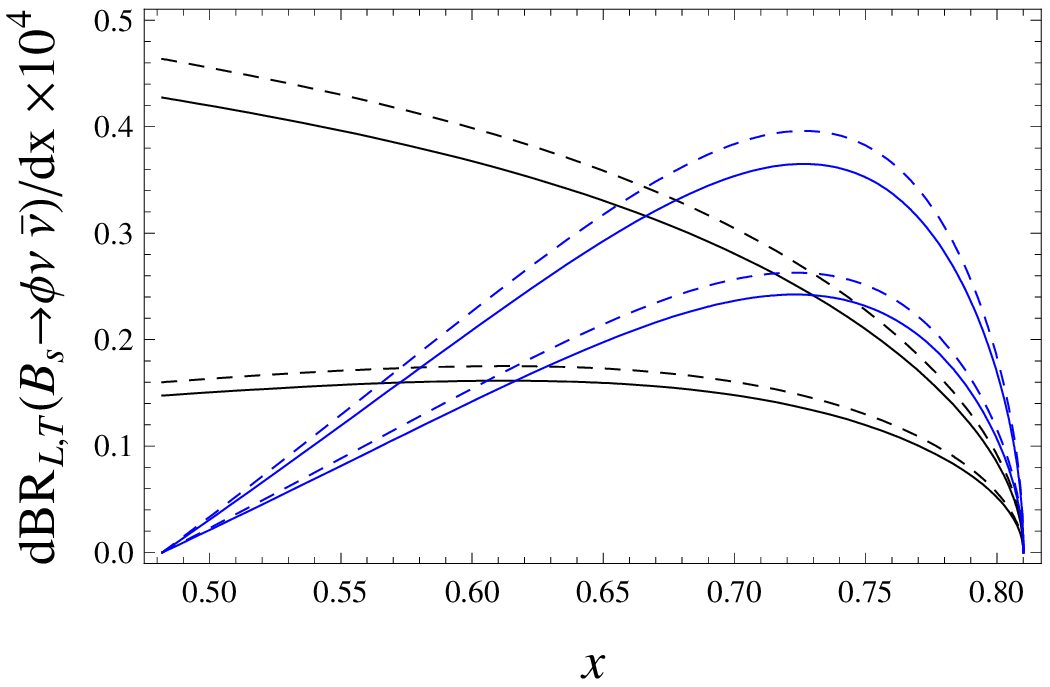}\\
 \includegraphics[width=0.37\textwidth] {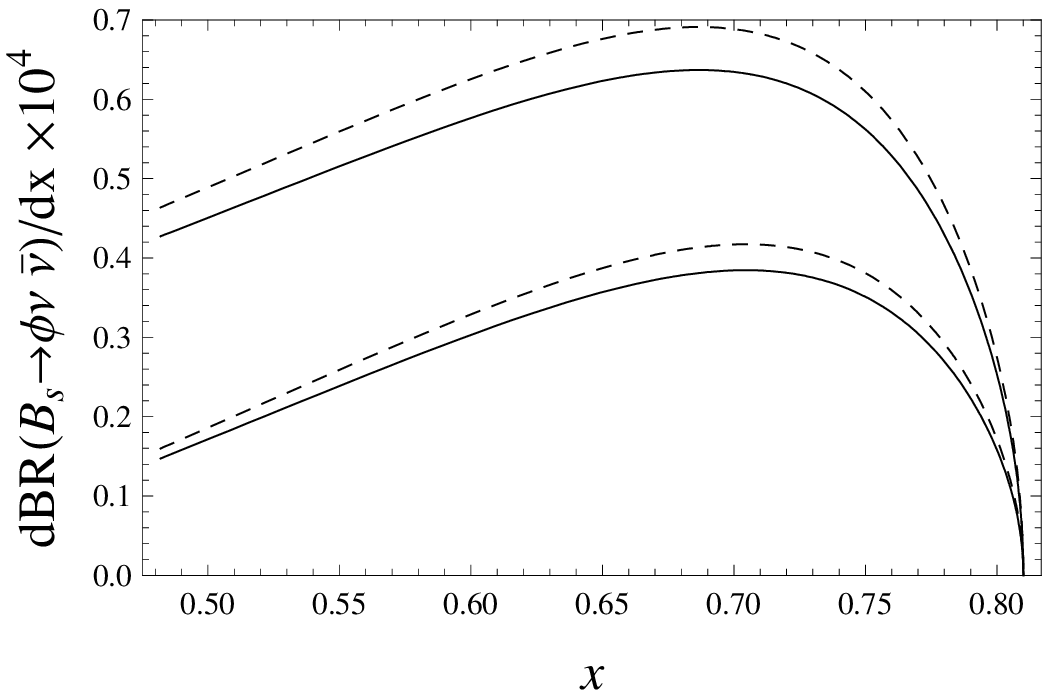}
 \end{center}
\caption{\baselineskip=12pt    Missing energy distribution in  $B_s
\to \phi \nu {\bar \nu}$  (upper panel) for  longitudinally 
(left curves) and transversally polarized $\phi$ meson (right curves).
The sum over the three neutrino species is understood. The continuous lines bound the region
corresponding to SM, the dashed lines the region corresponding to
 $1/R=500$ GeV.
 In the lower panel the missing energy distribution for
unpolarized $\phi$ is depicted  (same notations).}
\label{spectrumLTnunu}
\end{figure}

The missing energy distributions for polarized and
unpolarized $\phi$  are depicted in Fig. \ref{spectrumLTnunu} for $1/R=500$ GeV and in the
Standard Model.  They are
obtained using  $B_s \to \phi$ form factors determined by 
LCSR  \cite{Ball:2004rg}:
\bea A_1(q^2)&=&{A_1(0) \over 1- {q^2 \over m^2_{A_1}}}
\nn \\
A_2(q^2)&=& {r_1^{A_2} \over  1- {q^2 \over m^2_{A_2}}}+
{r_2^{A_2} \over  \left(1- {q^2 \over m^2_{A_2}} \right)^2} \\
V(q^2)&=& {r_1^V \over  1- {q^2 \over m^2_R}}+ {r_2^V \over  1-
{q^2 \over m^2_V} } \nn \eea \noindent where $A_1(0)=0.311 \pm
0.030 $ and $m^2_{A_1}=36.54$ GeV$^2$; $r_1^{A_2}=-0.054$,
$r_2^{A_2}=0.288$, $A_2(0)=r_1^{A_2}+r_2^{A_2}=0.234 \pm 0.028 $,
$m^2_{A_2}=48.94$ GeV$^2$; $r_1^V=1.484$, $r_2^V=-1.049$,
$V(0)=r_1^V+r_2^V=0.434 \pm 0.035$, $m_R=5.32$ GeV, $m^2_V=39.52$
GeV$^2$.
The effect of  the extra dimension consists in 
 a systematic increase of   the various distributions
in the full range of missing energy;
however, the hadronic uncertainty needs to be substantially reduced in order to clearly disentangle
deviations from the Standard Model predictions which correspond to this  value of  $1/R$.

\begin{figure}[t]
\begin{center}
\vspace*{0.3cm}
 \includegraphics[width=0.39\textwidth] {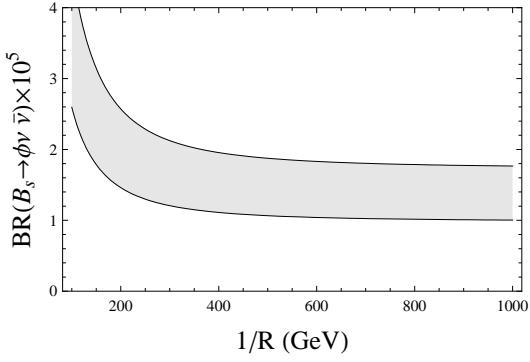}
\end{center}
\caption{\baselineskip=12pt ${\cal B}(B_s \to \phi \nu {\bar \nu})$
{\it vs} the compactification parameter  $1/R$ in the ACD model. } \vspace*{1.0cm} \label{brbsphinunu}
\end{figure}

In the SM  the branching ratio is predicted: 
 \be
 {\cal B}(B_s \to
\phi \nu \bar \nu)=(1.4 \pm 0.4) \times 10^{-5} \label{brbsnunusm}\ee 
therefore  this mode is  within the reach of  future experiments, at least as far as the number of produced  events is concerned, although  the observation of a  final state involving a neutrino-antineutrino pair is  a challenging task, as observed also in  \cite{Geng:2003su}.
The dependence of  ${\cal B}(B_s \to \phi \nu \bar \nu)$ on  $1/R$ is depicted in Fig. \ref{brbsphinunu}, 
where it is shown  that  the branching fraction increases for low values of
$1/R$:  for example,  at $1/R=300$ GeV there is a $ 23\%$ enhancement with
respect to  the SM expectation.  For larger values  of $1/R$ the Standard Model prediction is recovered,  the dependence on $1/R$ being  obscured  by  form factor uncertainties.

\section{$\Lambda_b \to \Lambda \gamma$ and $\Lambda_b \to \Lambda \nu \bar \nu$ decays}\label{sec:lambdab}

In  case of $\Lambda_b \to \Lambda$ transitions  the hadronic
 matrix elements of the operators  $O_7$ and $O_L$ in eqs. (\ref{hamilgamma})
and (\ref{hamilnu}) involve a larger number  of form factors. As a matter of fact, the various matrix
elements can be written as follows:

\bea &&<\Lambda(p^\prime,s^\prime)|{\bar
s}i \sigma_{\mu \nu} q^\nu b |\Lambda_b(p,s)> \nonumber  \\
&&= {\bar u}_\Lambda \big[ f_1^T(q^2) \gamma_\mu +i
f_2^T(q^2)\sigma_{\mu \nu} q^\nu+f_3^T(q^2) q_\mu \big]
u_{\Lambda_b}  \nonumber \\  \label{lambda1}   
\eea 
\bea 
&&<\Lambda(p^\prime,s^\prime)|{\bar s}i \sigma_{\mu \nu} q^\nu \gamma_5 b
|\Lambda_b(p,s)> \nonumber \\ &&= {\bar u}_\Lambda
\big[ g_1^T(q^2) \gamma_\mu \gamma_5  +i g_2^T(q^2)\sigma_{\mu
\nu} q^\nu \gamma_5 +g_3^T(q^2) q_\mu \gamma_5 \big] u_{\Lambda_b}
\nonumber \\ \label{lambda2}
\eea
\bea 
&&<\Lambda(p^\prime,s^\prime)|{\bar s} \gamma_{\mu}  b
|\Lambda_b(p,s)> \nonumber  \\ 
&&= {\bar u}_\Lambda
\big[ f_1(q^2) \gamma_\mu +i f_2(q^2)\sigma_{\mu \nu}
q^\nu+f_3(q^2) q_\mu \big] u_{\Lambda_b} \nonumber \\   \label{lambda3} 
 \eea
 \bea
&&<\Lambda(p^\prime,s^\prime)|{\bar s} \gamma_{\mu} \gamma_5 b
|\Lambda_b(p,s)> \nonumber \\
&&= {\bar u}_\Lambda
\big[ g_1(q^2) \gamma_\mu \gamma_5  +i g_2(q^2)\sigma_{\mu \nu}
q^\nu \gamma_5 +g_3(q^2) q_\mu \gamma_5 \big] u_{\Lambda_b}
\nonumber \,,  \\ \label{lambda4} \eea 
with $u_\Lambda$ and $u_{\Lambda_b}$  the
$\Lambda$  and $\Lambda_b$ spinors.

At present, a determination  of all the form factors in (\ref{lambda1})-(\ref{lambda4})
is not available. However, it is possible to  invoke
heavy quark symmetries for the hadronic matrix elements involving    an initial   spin=${1 \over 2}$ heavy baryon comprising a single heavy quark Q and a final   ${1 \over 2}$ light baryon;  the heavy quark symmetries    reduce to two the
number of independent form factors. As a matter of fact, in the infinite heavy quark limit $m_Q \to \infty$ and  for a generic Dirac matrix $\Gamma$ one can write \cite{Mannel:1990vg}: 
\bea &&<\Lambda(p^\prime,s^\prime)|{\bar s} \Gamma b
|\Lambda_b(p,s)> \nonumber\\ &&= {\bar
u}_\Lambda(p^\prime,s^\prime) \big\{ F_1(p^\prime \cdot v) + \spur{v} F_2  (p^\prime \cdot v)
\big\} \Gamma u_{\Lambda_b}(v,s) \nonumber \\  \label{hqrelations} \eea where 
$v={p \over M_{\Lambda_b}}$  is the $\Lambda_b$ four-velocity. 
The form factors $F_{1}$ and $F_{2}$ depend on $\dd{p^\prime \cdot v= {M^2_{\Lambda_b}+M^2_\Lambda-q^2 \over 2 M_{\Lambda_b}}}$ (for convenience we instead consider them as functions of
$q^2$ through this relation).
The expression (\ref{hqrelations}) for a generic matrix element 
not only shows  that the number of independent form factors is reduced to two, but  also that such form factors are
universal, since the same functions $F_1$ and 
$F_2$ describe both $\Lambda_b$, both $\Lambda_c$ decays into $\Lambda$,  envisaging the possibility of relating these two kind of  processes  if finite quark mass effects, in particular in the charm case, are small.  
The relations between the form factors in (\ref{lambda1})-(\ref{lambda4})
and the universal functions in   (\ref{hqrelations}):
\bea f_1 &=& g_1 = f_2^T=g_2^T=F_1 +{M_\Lambda \over
M_{\Lambda_b}} F_2 \nonumber \\
f_2 &=& g_2=f_3=g_3={F_2 \over M_{\Lambda_b}}\nonumber \\
f_1^T &=& g_1^T=q^2 {F_2 \over M_{\Lambda_b}}  \label{ffrelations} \\
f_3^T &=& -(M_{\Lambda_b}- M_\Lambda){F_2 \over M_{\Lambda_b}}\nonumber \\
g_3^T &=& (M_{\Lambda_b}+ M_\Lambda){F_2 \over M_{\Lambda_b}} 
\nonumber \eea
 are strictly valid  at  momentum transfer close to the maximum value
$q^2 \simeq q^2_{max}=(M_{\Lambda_b}-M_\Lambda)^2$. 
 However, we extend their validity  to  the whole phase
space, an assumption which introduces a   model dependence in the predictions. 

A determination of  $F_1$ and  $F_2$ has been  obtained by
three-point QCD sum rules in the $m_Q \to \infty$ limit  \cite{Huang:1998ek}. In  the
following  we  use the expressions for the
functions $F_1$, $F_2$ obtained by  updating some of the parameters
used  in \cite{Huang:1998ek}. In particular,  we use the PDG value  of the  $\Lambda_b$ mass: 
 $M_{\Lambda_b}=5.624\pm0.009$ GeV  \cite{PDG}.  Moreover, we fix   the
mass difference $\Delta_{\Lambda_b}=M_{\Lambda_b} -m_b$,  together with a constant $f_{\Lambda_b}$ 
parameterizing a vacuum-current  $\Lambda_b$ matrix element,  to the values
computed   in the infinite heavy quark mass limit in  \cite{Colangelo:1995qp}:
 $\Delta_{\Lambda_b}=0.9\pm 0.1$ GeV and $f_{\Lambda_b}=(2.9 \pm 0.5)\times 10^{-2}$
 GeV$^3$.
 Using these inputs, the obtained form factors  $F_{1,2}$  can be parameterized  by the expressions:
 \be
 F_{1,2}(q^2)={F_{1,2}(0) \over 1+a_{1,2} \, q^2+b_{1,2} \, q^4} \label{fit-ff-huang}\ee
with 
$F_1(0) = 0.322 \pm 0.015$,     $a_1=-0.0187$ GeV$^{-2}$,    $b_1=-1.6\times 10^{-4}$ GeV$^{-4}$,  and
$F_2(0) =-0.054 \pm  0.020$,    $a_2=-0.069$ GeV$^{-2}$,   $b_2=1.5\times 10^{-3}$ GeV$^{-4}$.

A few remarks are in order. First, the $q^2$ (or $p^\prime \cdot v$) dependence of the two form factors turns out to be different,   at odds with what has been assumed in various analyses  where the same
$q^2$ dependence for $F_1$ and $F_2$ is argued \cite{Korner:1991ph}.
Second, while $F_1$ and $F_2$ are monotonic in $q^2$, their dependence is different from the
simple or multiple pole behaviour assumed in other analyses, with pole mass fixed by vector meson dominance
(VMD) arguments.
Finally, $F_2$ is different from zero. This is noticeable, even though in  various decay rates the terms
involving $F_2$  appear together with the suppressing  factor $1/M_{\Lambda_b}$.  An analysis of the helicity amplitudes in  $\Lambda_c \to \Lambda \ell \nu$
carried out  by the CLEO collaboration measuring various angular distributions
in this decay process   demonstrated  that  $F_2$ is not vanishing \cite{Hinson:2004pj}.  The result,  based on the assumption 
 that the heavy quark limit can be applied in the charm case and on the hypothesis the both $F_1$ and $F_2$ have the same  dipolar $q^2$ dependence, is:
${F_{2}(q^2) \over F_{1}(q^2)}= -0.35 \pm0.04\pm0.04$ if the pole mass is fixed to  $M_{pole}=M_{D^*_s}$, or ${F_{2}(q^2) \over F_{1}(q^2)}= -0.31 \pm0.05\pm0.04$ in correspondence to a fitted  pole mass $M_{pole}=2.21\pm0.08\pm0.14$ GeV. 

The idea of using measurements of observables in  semileptonic $\Lambda_c \to \Lambda \ell \nu$ transitions to determine the universal form factors $F_1$ and $F_2$, with the aim of employing  them  
to describe   several $\Lambda_b$ transitions, 
   was proposed in \cite{Mannel:1997xy}.  However, in such a proposal the $q^2$ dependence
of the two form factors must be assumed.   If  $F_{1,2}$ have both a dipolar $q^2$ dependence, using
the ratio $\dd {F_{2} \over F_{1}}$ determined by the CLEO collaboration and the experimental value of
 ${\cal B}(\Lambda_c \to \Lambda \ell \nu)$, one gets:   
$F_{1}(q^2_{max})=1.1\pm0.2$.  Following  the strategy proposed in \cite{Mannel:1997xy}
it is necessary to
extrapolate $F_1$ to the full kinematical range allowed in $\Lambda_b \to \Lambda$ transitions
by the assumed $q^2$ dependence, and apply the result  to $\Lambda_b \to \Lambda \gamma$ and  $\Lambda_b \to \Lambda \nu \bar \nu$. We refrain from applying such a procedure, since the momentum range where the extrapolation is required is very wide and,
more importantly,  the assumption that finite charm mass effects are negligible in case of $\Lambda_c$ should be justified.  

It is worth mentioning that in the $m_b \to \infty$ limit and in the large recoil regime for the light
hadron ($q^2 \simeq 0$ or $E=v \cdot  p^\prime \to \infty$), the relation
\be
{F_2(0)\over F_1(0)} =- {M_\Lambda \over 2 E} \label{SCET}
\ee
can be derived using the Large Energy Effective Theory/SCET framework
at the leading order in the $\dd 1 \over m_b$ and $\dd 1 \over E$ expansion
 \cite{hiller}.  Direct calculations
of  $F_{1,2}$ in this limit have not been done, yet;   the  model   (\ref{fit-ff-huang})
gives a ratio of form factors compatible with  (\ref{SCET}).

Admittedly,  the knowledge of  $\Lambda_b$ form factors deserves a
substantial  improvement;  in the meanwhile, we use in our analysis the form factors in (\ref{fit-ff-huang}) stressing that the uncertainties we attach to the various predictions  only take into account  the errors  of  the parameters of the expressions used for   $F_1$ and $F_2$.


\subsection{$\Lambda_b \to \Lambda \gamma$}

\begin{figure}[t]
\begin{center}
 \includegraphics[width=0.40\textwidth] {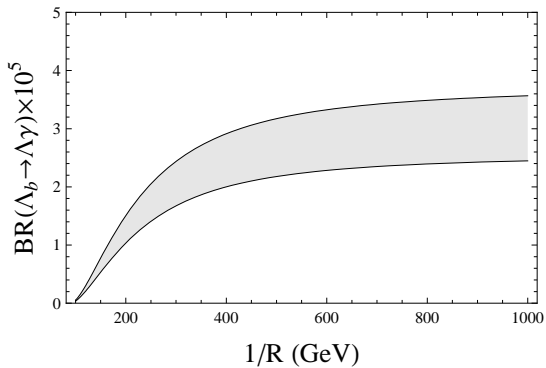}
\end{center}
\caption{\baselineskip=12pt ${\cal B}(\Lambda_b \to \Lambda \gamma)$ {\it vs} 
 $1/R$.   The uncertainty shown by  the dark band is mainly due to the errors on the form 
 factors of the  model (\ref{fit-ff-huang}).}
\vspace*{1.0cm} \label{brllgammahuang}
\end{figure}

The expression of the $\Lambda_b \to \Lambda \gamma$ decay rate: 
\bea 
&&
\Gamma(\Lambda_b \to \Lambda \gamma) = {\alpha(0) G_F^2 |V_{tb}
V_{ts}^*|^2 m_b^2 \over 32 \pi^4} |c_7^{eff}|^2  M_{\Lambda_b}^3 \nonumber  \\
&& \times  \left(1-{M_\Lambda^2 \over M_{\Lambda_b}^2} \right)^3 
\left( F_1(0)+{M_\Lambda \over M_{\Lambda_b}} F_2(0) \right)^2 \,\,\, \nn \\ \label{rategamma} \eea 
is useful to relate also this mode to the observed $B_d \to K^{*0} \gamma$ transition:
\bea
&& {\cal B}(\Lambda_b \to \Lambda \gamma)= \left ({F_1(0) \over T_1^{B_d \to K^{*0}}(0)}\right)^2 \left(1+{M_\Lambda \over M_{\Lambda_b}} {F_2(0) \over F_1(0)} \right)^2 \nn \\
 && \times \left( {M_{B_d} \over M_{\Lambda_b}} \right)^3 \left( {M_{\Lambda_b}^2-M_\Lambda^2 \over M_{B_d}^2-M_{K^{*0}}^2}\right)^3 {\tau_{\Lambda_b} \over  4 \tau_{B_d}} {\cal B}(B_d \to K^{*0} \gamma) \,\,\, . \nn \\ \label{Lbratio}
\eea
Using
$\tau_{\Lambda_b}=(1.23 \pm0.074) \times 10^{-12} \,s$ \cite{PDG},
together with the ratio of form factors  ${F_2(0) \over F_1(0)}=-0.17\pm0.06$
and 
$\dd{F_1(0) \over T_1^{B_d \to K^{*0}}(0)}=1.9\pm 0.2$
obtained from  (\ref{fit-ff-huang}) and the form factor $T_1$ in \cite{Ball:2004rg},
 we predict:  ${\cal B}(\Lambda_b \to \Lambda \gamma)=(3.4 \pm 0.7) \times 10^{-5}$.
 The same result with a larger uncertainty comes from using $T_1$ determined in \cite{sant}.
 Using (\ref{rategamma})  and the form factors (\ref{fit-ff-huang}), we get:
\be
 {\cal B}(\Lambda_b \to \Lambda \gamma)=(3.1 \pm 0.6) \times 10^{-5} \,\,\, .
\label{brllgammasm}\ee
The result suggests that this process  is within the reach of  LHC experiments. 
As for  the effect of the extra dimension in modifying the decay rate, in Fig. \ref{brllgammahuang} we show how  ${\cal B}(\Lambda_b \to
\Lambda \gamma)$ depends on $1/R$. Analogously to  $B\to K^* \gamma$ and $B_s\to \phi \gamma$
transitions, the branching fraction is suppressed for low values
of $1/R$, being $30\%$ smaller for $1/R=300$ GeV. 

\vspace*{0.5cm}

\subsection{$\Lambda_b \to \Lambda \nu \bar \nu$}
\begin{figure}[th]
\begin{center}
\includegraphics[width=0.37\textwidth] {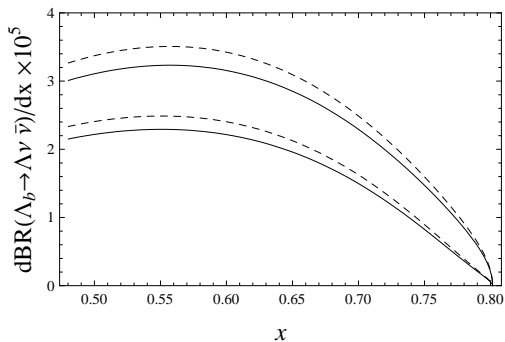}
\end{center}
\caption{\baselineskip=12pt  Missing energy distribution for
$\Lambda_b \to \Lambda \nu \bar \nu$ in the Standard Model (continuous lines) and
for $1/R=500$ GeV (dashed lines).
 } \vspace*{1.0cm} \label{spectrumllnunu}
\end{figure}

\begin{figure}[t]
\begin{center}
 \includegraphics[width=0.40\textwidth] {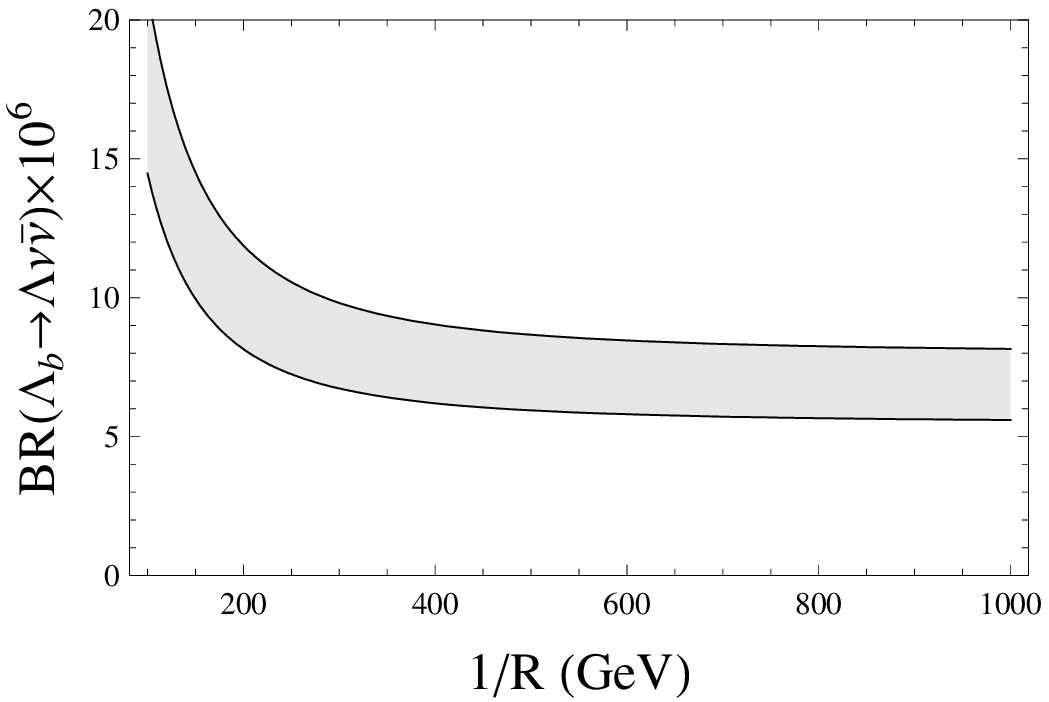}
\end{center}
\caption{\baselineskip=12pt ${\cal B}(\Lambda_b \to \Lambda \nu \bar
\nu)$ versus $1/R$. 
} \vspace*{1.0cm} \label{brllnunuhuang}
\end{figure}

Even for this mode  it is useful to consider
the missing energy distribution in  the variable
$x=E_{miss}/M_{\Lambda_b}$ with $E_{miss}$  the energy of the
neutrino - antineutrino pair:
\bea
&&{d \Gamma \over dx }= {|c_L|^2| \vec p^\prime| \over 4 \pi^3 } \nonumber \\ && 
\Big\{ \Big[ \big( M_{\Lambda_b}^2-M_{\Lambda}^2
\big)^2+q^2 \big( M_{\Lambda_b}^2+M_{\Lambda}^2-2 q^2 \big) \Big] \nonumber \\ && 
\times \Big[F_1+ {M_{\Lambda} \over M_{\Lambda_b}} F_2 \Big]^2 \nonumber \\
&& + \Big[2 q^2 \big( M_{\Lambda_b}^2-M_{\Lambda}^2 \big)^2 -q^4
\big( M_{\Lambda_b}^2+M_{\Lambda}^2+q^2 \big) \Big] \Big({F_2 \over M_{\Lambda_b}}\Big)^2 \nonumber \\
&& + 6 M_{\Lambda} q^2 \big( M_{\Lambda_b}^2-M_{\Lambda}^2+q^2 \big)
\Big[F_1+ {M_{\Lambda} \over M_{\Lambda_b}} F_2 \Big]{F_2 \over
M_{\Lambda_b}} \Big\}\,. \nonumber \\ 
\label{dgammalnunubar} \eea
 In Fig. \ref{spectrumllnunu} we plot such a distribution in the
 SM and for  $1/R=500$ GeV,  showing the differences corresponding to such a value of  the
 compactification radius. 
 The  branching ratio, which in SM is expected to be  \cite{Chen:2000mr}:
  \be
 {\cal B}(\Lambda_b \to \Lambda \nu \bar \nu)=(6.7 \pm 1.3) \times 10^{-6}
\label{brllnunusmhuang}\ee 
depends on $1/R$ as shown in Fig.  \ref{brllnunuhuang}.
When $1/R$ decreases the branching ratio increases
 of about  $20\%$  for  $1/R\simeq 300$ GeV.

%
%
%
%

\section{Conclusions} \label{sec:concl}

We have studied how  a single universal extra dimension
could have an impact on  several  loop induced $B_s$ and $\Lambda_b$ decays. The analysis of processes involving these
two particles will be among the main topics in the investigations at the hadron colliders, especially at LHC. Since a few of the modes we have considered are  difficult to reconstruct in a "hostile" environment represented by  hadron collisions, we believe that the predictions we have
worked out  are useful to elaborate the measurement strategies.  From the theoretical point of view,
we have found that hadronic uncertainties due to the form factors in exclusive decays are not large
 in case of $B_s$, where $SU(3)_F$ symmetry is also useful to  exploit  other
measurements carried out at the $B$ factories.
As for $\Lambda_b$, the situation is more uncertain. Calculations made in the infinite heavy quark limit should be corroborated by the analysis of finite heavy quark mass effects, and the various dependences on the momentum transfer should be confirmed. Such analyses deserve a dedicated effort. 

\vspace*{0.5cm}
\begin{center}
{\bf Acknowledgments} \end{center}
F.D.F. thanks M. Misiak for very useful discussions. This work was supported in
part by the EU contract No. MRTN-CT-2006-035482, "FLAVIAnet".

\clearpage


\end{document}